\documentclass[12pt,reqno]{article}
\usepackage{amsmath,hyperref,amssymb}
\usepackage{graphicx,url}

\allowdisplaybreaks \numberwithin{equation}{section}

\DeclareSymbolFont{AMSa}{U}{msa}{m}{n}
\DeclareSymbolFont{AMSb}{U}{msb}{m}{n}
\DeclareMathSymbol{\fieldR}{\mathalpha}{AMSb}{"52}

\begin{document}

\begin{flushright} \small
ITP--UU--11/46 \\ SPIN--11/36
\end{flushright}
\bigskip
\begin{center}
 {\large\bfseries On BPS bounds in D=4 N=2 gauged supergravity II:\\ general matter couplings and black hole masses}\\[5mm]
Kiril Hristov\\[3mm]
 {\small\slshape
 Institute for Theoretical Physics \emph{and} Spinoza Institute, \\
 Utrecht University, 3508 TD Utrecht, The Netherlands \\
\medskip
 Faculty of Physics, Sofia University, Sofia 1164, Bulgaria\\
\medskip
 {\upshape\ttfamily K.P.Hristov@uu.nl}\\[3mm]}
\end{center}
\vspace{5mm} \hrule\bigskip \centerline{\bfseries Abstract}
We continue the analysis of BPS bounds started in \cite{Hristov:2011ye}, extending it to the full class of $N=2$ gauged supergravity theories with arbitrary vector and hypermultiplets. We derive the general form of the asymptotic charges for asymptotically flat (M$_4$), anti-de Sitter (AdS$_4$), and magnetic anti-de Sitter (mAdS$_4$) spacetimes. Some particular examples from black hole physics are given to explicitly demonstrate how AdS and mAdS masses differ when solutions with non-trivial scalar profiles are considered.
\bigskip
\hrule\bigskip

\section{Introduction and general results}\label{sect:BPS_bounds}
This paper is a continuation of the work of \cite{Hristov:2011ye} and aims at a derivation of the BPS bounds for solutions of gauged $D=4$ $N=2$ supergravity with vector and hypermultiplets. We briefly recall that in \cite{Hristov:2011ye} a method was developed for explicit evaluation of BPS bounds for solutions in supergravity, based on their asymptotic Killing spinors. The main results were the derivation of the asymptotic charges in minimal gauged supergravity for asymptotically AdS and magnetic AdS solutions, which differ by their magnetic charge. For stationary solutions, the BPS bound in AdS with vanishing magnetic charge $Q_m = 0$ is found to be
\begin{align}\label{BPS_ads}
    M \geq |Q_e| + g |\vec{J}|\ ,
\end{align}
with $M$ the mass, $Q_e$ the electric charge, $\vec{J}$ the angular momentum of the given solution, and $g$ the gauge coupling that is related to the cosmological constant. For asymptotically mAdS solutions on the other hand, the BPS bound is
\begin{align}\label{BPS_mads}
    M \geq 0\ ,
\end{align}
with magnetic charge $Q_m = \pm 1/(2g)$.

As we show in the present work, the superalgebra structure does not change when considering more general matter
couplings in the theory. Thus, \eqref{BPS_ads} and \eqref{BPS_mads} continue to hold. However, the explicit
definition of the asymptotic charges ($M, Q_e,$ etc.) of a given solution depends directly on the field content.
We first derive the form of the supersymmetry anticommutator for all possible solutions of gauged supergravity
with vectors and hypers. Then we focus on the special cases of Minkowski, AdS, and mAdS asymptotics where we
evaluate the anticommutator explicitly. These calculations show that the hypermultiplets do not produce
additional central charges in the superalgebra. We are also able to formulate renormalized expressions for the
mass in AdS and mAdS. Our results in AdS are in exact agreement with the techniques of holographic
renormalization \cite{Kostas}. On the other hand, the mAdS mass takes a different form and in some examples
leads to qualitatively different results that have no analog in previous literature.

We consider the most general (two-derivative) electrically\footnote{Although explicitly concentrating on
electric gaugings here, the results will hold for more general theories with electromagnetic gauging such as the
ones described in \cite{bernard_maaike}. This is due to the fact that electromagnetic duality rotates the
symplectic frame of the general lagrangian of \cite{bernard_maaike} and one can always find a purely electric
frame, where our results hold exactly. Since the spectrum of the theories remains invariant under symplectic
transformations, our results generalize trivially.} gauged $D=4$ $N=2$ supergravity, following strictly the
conventions of \cite{Hristov:2009uj} (that are mostly the same as in \cite{ABCDF}). For further background
material on $N=2$ supergravity, see e.g.\ \cite{deWit,DeWit:1984px,D'Auria:1990fj}. The standard $N=2$ graviton
multiplet (graviton $g_{\mu \nu}$, graviphoton $A_{\mu}^g$ and two gravitinos) is coupled with $n_V$ vector
multiplets ($n_V$ complex scalars $z^i$, $n_V$ vectors $A_{\mu}^i$ and $2 n_V$ gauginos)\footnote{In the
lagrangian the graviphoton $A_{\mu}^g$ and vector fields $A_{\mu}^i$ mix between each other and appear as vector
fields $A_{\mu}^{\Lambda}$, $\Lambda = 0,...,n_V$, with corresponding field strengths $F_{\mu \nu}^{\Lambda}$.}
and $n_H$ hypermultiplets ($4 n_H$ real scalars $q^u$ and $2 n_H$ hyperinos). The bosonic part of the lagrangian
is
\begin{eqnarray}\label{lagr}
{\cal L}&=&\frac{1}{2}R(g)+g_{i\bar \jmath}(z,\bar{z})\nabla^\mu z^i
\nabla_\mu
{\bar z}^{\bar \jmath} + h_{uv}(q)\nabla^\mu q^u \nabla_\mu q^v + I_{\Lambda\Sigma}(z,\bar{z})F_{\mu\nu}^{\Lambda}F^{\Sigma\,\mu\nu}\\
&+&\frac{1}{2}R_{\Lambda\Sigma}(z,\bar{z})\epsilon^{\mu\nu\rho\sigma}
F_{\mu\nu}^{\Lambda}F^{\Sigma}_{\rho\sigma}-\frac{4}{3}g\,c_{\Lambda,\Sigma\Pi}\,\epsilon^{\mu\nu\rho\sigma}A_\mu^\Lambda
A_\nu^\Sigma \left( \partial_\rho
A_\sigma^\Pi-\frac{3}{8}f_{\Omega\Gamma}{}^\Pi A_\rho^\Omega
A_\sigma^\Gamma \right) -V(z,\bar z,q)\ ,\nonumber
\end{eqnarray}
with scalar potential
\begin{equation}\label{pot2}
V=g^2\Big[ (g_{i\bar \jmath}k^i_\Lambda k^{\bar \jmath}_\Sigma + 4
h_{uv}k^u_\Lambda k^v_\Sigma){\bar L}^\Lambda L^\Sigma + (g^{i\bar
\jmath}f_i^\Lambda {\bar f}_{\bar \jmath}^\Sigma -3{\bar
L}^\Lambda L^\Sigma)P^x_\Lambda P^x_{\Sigma}\Big]\ .
\end{equation}
Most of the above quantities and the supersymmetry transformations will not be important for our purposes here so we leave the more technical introduction to the full lagrangian to appendix \ref{app:A}. The quantities of relevance for the derivation of the BPS bound will be introduced shortly when needed. As described in detail in \cite{Hristov:2011ye}, one in principle needs to consider the full lagrangian (or just upto second order terms in fermions when eventually setting fermions to zero) in order to derive the expression for the supercharges. Alternatively, one can fix the right form of the supercharges from the supersymmetry variations. From our knowledge of the minimal case \cite{Hristov:2011ye} and with the help of the susy variations we can derive explicitly the supercharge, as done in appendix \ref{app:B}. The original expression for the supercharge is somewhat lengthy and non-suggestive. However, using the equations of motion for the gravitinos we can cast the supercharge into a much simpler form as a surface integral (see the appendix for the technical details).

The important quantity for our purposes here is the Dirac bracket of two supercharges. It can be derived from the supercharge \eqref{susy_charge} and takes the remarkably simple form
\begin{align}\label{basic_susy_anticommutator_general}
\{\mathcal{Q},\mathcal{Q} \}  = \oint_{\partial V} {\rm d}\Sigma_{\mu \nu}(\epsilon^{\mu \nu \rho \sigma}
\overline{\varepsilon}^A \gamma_{\rho} \widetilde{\mathcal{D}}_{\sigma} \varepsilon_A - \epsilon^{\mu \nu \rho \sigma}
\overline{\varepsilon}_A \gamma_{\rho} \widetilde{\mathcal{D}}_{\sigma} \varepsilon^A)\ ,
\end{align}
where
\begin{align}\label{supercov_der}
\widetilde{\mathcal{D}}_{\mu} \varepsilon_A = (\partial_\mu -\frac{1}{4}{\cal
\omega}_\mu^{ab}\gamma_{ab})\varepsilon_A+\frac{i}{2} A_\mu
\varepsilon_A+\omega_{\mu\,A}{}^B\varepsilon_B +
T^-_{\mu\nu}\gamma^\nu
\epsilon_{AB}\varepsilon^B+igS_{AB}\gamma_\mu\varepsilon^B\ .
\end{align}
Here ${\cal \omega}_\mu^{ab}$ is the spin connection, $A_{\mu}$ is the gauged $U(1)$ K\"{a}hler connection,
\begin{equation}\label{gaugedU1}
A_\mu\equiv -\frac{i}{2}\Big(\partial_i{\cal K}\nabla_\mu
z^i-\partial_{\bar\iota}{\cal K}\nabla_\mu{\bar z}^{\bar
\iota}\Big)\ ,
\end{equation}
and $\omega_{\mu\,A}{}^B$ is the gauged $Sp(1)$ connection of the quaternion-K\"{a}hler manifold,
\begin{equation}\label{gauged-Sp1}
 \omega_{\mu \,A}{}^B\equiv\partial_\mu q^u \omega_{u\,A}{}^B+gA_\mu^\Lambda P^x_\Lambda (\sigma^x)_A{}^B\ .
 \end{equation}
 The quantity $T^-_{\mu\nu}$ is the anti-selfdual part of the graviphoton field strength,
 \begin{align}
T^-_{\mu\nu}\equiv 2iF^{\Lambda\,-}_{\mu\nu}({\rm I}_{\Lambda\Sigma}) L^\Sigma \ ,
\end{align} and
 \begin{align}
 S_{AB} \equiv \frac{i}{2}(\sigma^x)_{AB} P^x_\Lambda L^\Lambda
 \end{align} is the gravitino mass matrix (see App.\ \ref{app:A} and \cite{ABCDF} for more details about special and quaternion K\"{a}hler geometry). Eq.\ \eqref{basic_susy_anticommutator_general} is the main general result of this paper. It can be explicitly evaluated on every spacetime that has an asymptotic Killing spinor\footnote{Everywhere in this paper the solutions of $\widetilde{\mathcal{D}}_{\mu} \varepsilon_A = 0$ are referred to as Killing spinors. Each independent Killing spinor signifies the existence of a preserved fermionic isometry, i.e.\ supersymmetry.}.

Compared with the corresponding expression in the minimal case \cite{Hristov:2011ye}, \eqref{basic_susy_anticommutator_general} is just a straightforward generalization. A priori, one could expect some more radical changes due to the presence of vector and hypermultiplets, but this is not the case. We already see that the main conclusions of \cite{Hristov:2011ye} remain the same, with the difference that the definition of the asymptotic charges will generalize to accommodate for the possibility of non-constant scalars\footnote{Note that for a solution with constant scalars (both in the vector and in the hypermultiplet sector) \eqref{basic_susy_anticommutator_general} is equivalent with the result for the minimal case. Thus, the only difference between the asymptotic charges in minimal and non-minimal supergravity lies in the possibility for non-constant scalar profiles.}. In order to give more precise statements, we need to plug in the explicit Killing spinors of interest in the general Dirac bracket \eqref{basic_susy_anticommutator_general} as described in section 3 of \cite{Hristov:2011ye}.

 In the following sections we consider more carefully the cases of Minkowski, AdS$_4$, and mAdS$_4$ asymptotics, paying special attention to the asymptotic
charges in stationary solutions. In each of the cases we give an explicit example from the study of black holes
as an application of our results. Somewhat surprisingly, we are able to find a very simple unified formula for
the mass of supersymmetric black hole spacetimes in all three cases. This also leads to a better conceptual
understanding of the difference in the mass in AdS and mAdS spacetimes. We conclude with some remarks on the
connection of our results to alternative approaches in literature and mention other potential uses of our
method.

 \section{Asymptotically flat solutions}\label{sect:flat}
 \subsection{General analysis}
 Here we will be interested in the superalgebra and asymptotic charges of Minkowski spacetime. In the context of electrically gauged supergravity
with vector and hypermultiplets the necessary conditions for a Minkowski vacuum were derived in \cite{Hristov:2009uj},
\begin{align}
k^i_\Lambda \bar{L}^\Lambda =0\ , \qquad \tilde k^u_\Lambda L^\Lambda &=0\ , \qquad
P^x_\Lambda =0\ ,
\end{align}
together with constant scalars, vanishing field strengths and flat $\mathbb{R}^{1,3}$ metric. These are now the conditions that asymptotically flat solutions will have to satisfy as $r\rightarrow \infty$ (we always work in spherical coordinates as in \cite{Hristov:2011ye}).

The Majorana Killing spinors of Minkowski in spherical coordinates are
\begin{equation}\label{Killing_spinors-Minkowski}
    \tilde{\epsilon}^{1,2}_{M} = e^{-\frac{1}{2} \theta \gamma_{1 2}} e^{-\frac{1}{2}
    \varphi \gamma_{2 3}}\ \tilde{\epsilon}^{1,2}_0\ ,
\end{equation}
where $\tilde{\epsilon}^{1,2}_0$ are two arbitrary and linearly independent constant Majorana spinors. We will use the notation $\tilde{\epsilon}^A$ for Majorana spinors and $\varepsilon_A, \varepsilon^A$ for the positive/negative chirality Weyl spinors that are used in our notation. The chiral spinors are related to the Majorana ones through
\begin{align}\label{spinor_relation}
    \varepsilon_A \equiv \frac{1+\gamma_5}{2}\ \tilde{\epsilon}^A\ , \qquad \varepsilon^A \equiv \frac{1-\gamma_5}{2}\ \tilde{\epsilon}^A\ , \qquad (\varepsilon_A)^* = \varepsilon^A\ .
\end{align}

Having the Killing spinors we can now in principle plug \eqref{Killing_spinors-Minkowski} in \eqref{basic_susy_anticommutator_general} and derive the supercharge anticommutator directly. Of course, we already know the general answer from the Poincar\'{e} superalgebra,
\begin{align}\label{poincare_algebra}
\{ Q^{A \alpha},Q^{B \beta} \} =  \delta^{AB}
(i \gamma^{M}C^{-1})^{\alpha \beta} P_{M} - \epsilon^{AB} (({\rm Re}\mathcal{Z} + i \gamma^5 {\rm Im} \mathcal{Z})(C^{-1}))^{\alpha\beta}\ ,
\end{align}
where $C$ is the charge conjugation matrix, $P_M$ is the momentum operator, and $\mathcal{Z}$ is the complex central extension of the superalgebra. The explicit eigenvalues of the operators $P_M$ and $\mathcal{Z}$ for any asymptotically flat solution can be computed now from \eqref{basic_susy_anticommutator_general}. The additional $U(1)$ and $Sp(1)$ connections in \eqref{basic_susy_anticommutator_general} from the matter multiplets can potentially lead to contributions to the supersymmetry anticommutator that are not of the type \eqref{poincare_algebra}. Since we know that Minkowski asymptotics will necessarily lead to the Poincar\'{e} superalgebra it follows that these additional connections must fall off fast enough so that they do not contribute. \eqref{poincare_algebra} can in fact be taken as a definition for asymptotically flat spacetimes. In practice, the condition for the fall off of the connections will be equivalent with imposing the metric to approach Minkowski space. This will be illustrated more clearly with an explicit example.

In the next subsection we give the explicit expressions for $P_0, \mathcal{Z}$ in \eqref{poincare_algebra} for the stationary case, but one can straightforwardly derive the asymptotic charges in full generality if needed.

 \subsection{Stationary solutions}
 For stationary solutions we find that the supersymmetry anticommutator takes the following form\footnote{We rescale the central charges for convenience.}:
 \begin{align}\label{susy_commut_Mink}
    \{ Q^{A \alpha}, Q^{B \beta} \} = \delta^{A B} 8 \pi M (i \gamma^0 \mathcal{C}^{-1})^{\alpha \beta} - \epsilon^{A B} 8 \pi (({\rm Re}\mathcal{Z} + i \gamma^5 {\rm Im} \mathcal{Z}) (\mathcal{C}^{-1}))^{\alpha \beta}\ ,
 \end{align}
  where the complex central charge is given by
  \begin{align}\label{central_charge}
    \mathcal{Z} = \frac{1}{4\pi} \lim_{r \rightarrow \infty} \oint_{S^2} T^- = \lim_{r \rightarrow \infty}  \left( L^{\Lambda} q_{\Lambda} - M_{\Lambda} p^{\Lambda} \right)\ ,
  \end{align} as derived in detail in \cite{hep-th/9509160}\footnote{Note that the charges $q_{\Lambda}$ and $p^{\Lambda}$ in \eqref{central_charge} are the standard electric and magnetic charges as commonly defined in literature. The electric charges come from the dual field strengths $G_{\Lambda \mu \nu} \equiv i \epsilon_{\mu \nu \rho \sigma} \frac{\delta \mathcal{L}}{\delta F^{\Lambda}_{\rho \sigma}}$. See e.g. \cite{BLS,arXiv:1005.3650} for more details.}. The derivation of the central charge from \eqref{basic_susy_anticommutator_general} is a bit subtle and uses the fact that $\widetilde{\mathcal{D}}_{\mu} \varepsilon_A$ contains a $T_{\mu \nu}^-$ term, while $\widetilde{\mathcal{D}}_{\mu} \varepsilon^A$ contains $T_{\mu \nu}^+$. This eventually leads to $\int \left(T^- (1+ \gamma_5) + T^+ (1-\gamma_5) \right) \sim {\rm Re}\mathcal{Z} + i \gamma^5 {\rm Im} \mathcal{Z}$. This calculation picks out the electric and magnetic charge carried by the graviphoton, which explicitly depend on the asymptotic values of the vector multiplet scalars.

 The mass, on the other hand, remains unaffected by scalars,
 \begin{align}\label{mass_Mink}
    M = \frac{1}{8\pi}\lim_{r\rightarrow\infty} \oint {\rm d}\Sigma_{t r} \bigg( e^t_{[ 0} e^r_1 e^{\theta}_{2]} + \sin \theta\ e^t_{[ 0} e^r_1 e^{\varphi}_{3]} - (\omega_{\theta}^{a b} e^t_{[ 0} e^r_a e^{\theta}_{b]} + \omega_{\varphi}^{a b} e^t_{[ 0} e^r_a e^{\varphi}_{b]} ) \bigg)\ ,
 \end{align} just as in the minimal case. The vielbein and spin connection in the above formula can belong to any stationary asymptotically Minkowski solution of interest, explicit examples of such configurations can be found in the next subsection.

 The BPS bound, as always for stationary asymptotically flat solutions, is
 \begin{align}\label{bound_Mink}
    M \geq |\mathcal{Z}|\ .
 \end{align}
Note that the hypermultiplet sector seems to be completely decoupled from the above calculations since the hypers do not influence the asymptotic charges. This suggests that the stabilization of the hypers at a particular point in moduli space as described in \cite{arXiv:1005.3650} might be the generic situation in this case.

\subsection{Black hole example}
Example of asymptotically flat stationary solutions to apply the above formulas are hardly needed since these have been very well understood. As a standard example we can just briefly glance through the single-centered supersymmetric black holes of \cite{BLS}. First we take the most standard case of a static black hole as a warm up for the static examples in AdS and mAdS. We then also explain the case of a rotating BPS saturated Kerr-Newman metric, which provides a non-trivial test of the BPS bound \eqref{bound_Mink}.

The solutions of \cite{BLS} in ungauged supergravity allow for an arbitrary number of vector multiplets (and arbitrary hypermultiplets that decouple and will not be considered in what follows) with arbitrary charges $q_{\Lambda}, p^{\Lambda}$. The charges only need to satisfy a certain condition in order to make the metric static (see \cite{BLS} for more details). The metric and symplectic sections in spherical coordinates are
 \begin{align}\label{bls_solution}
 \begin{split}
 {\rm d} s^2 = e^{\mathcal{K}} ({\rm d} t^2 + \omega {\rm d} \varphi^2) - e^{-\mathcal{K}} {\rm d} r^2 -e^{-\mathcal{K}} r^2 {\rm d} \Omega_2^2\ ,\\
2\ {\rm Im} (X^{\Lambda}) = H^{\Lambda} = h^{\Lambda} + \frac{p^{\Lambda}}{r}, \qquad 2\ {\rm Im} (F_{\Lambda}) = H_{\Lambda} = h_{\Lambda} + \frac{q_{\Lambda}}{r}\ ,
 \end{split}
 \end{align}
 where $h^{\Lambda}, h_{\Lambda}$ are arbitrary constants that decide the asymptotic value of the scalars, usually chosen such that $e^{-\mathcal{K}}$ asymptotes exactly to $1$\footnote{One does not really need to stick to a particular choice for $h^{\Lambda}, h_{\Lambda}$. We can always perform a coordinate transformation to make sure that we have the correct asymptotics at $r \rightarrow \infty$. This has exactly the same effect.}. The rotation $\omega$ is present only when the K\"{a}hler connection \eqref{gaugedU1} is non-vanishing.

Let us consider as a first simple example the prepotential $F = - \frac{(X^1)^3}{X^0}$ with non-vanishing magnetic charge $p^0$ and electric charge $q_1$ (also non-vanishing $h^0, h_1$). This implies that $X^0 = \frac{i}{2} H^0, X^1 = \frac{1}{2} \sqrt{\frac{H^0 H_1}{3}}$ and $e^{-\mathcal{K}} = \frac{2}{3 \sqrt{3}} \sqrt{H^0 (H_1)^3}$. The $U(1)$ connection vanishes and therefore the metric is static, $\omega = 0$. To normalize the K\"{a}hler potential we choose $h^0 (h_1)^3 = \frac{27}{4}$ and find for the central charge
\begin{align}
    \mathcal{Z} = \frac{1}{4} \left(\frac{p_0}{h_0}+3 \frac{q^1}{h^1}\right)\ .
\end{align}
The mass can be calculated from \eqref{mass_Mink} with the metric \eqref{bls_solution} and spin connection $\omega_{\theta}^{12} = \frac{\omega_{\varphi}^{13}}{\sin \theta} = e^{\mathcal{K}/2} \partial_r (r e^{-\mathcal{K}/2})$ and becomes
\begin{align}\label{simple_mass_Mink}
    M = \lim_{r\rightarrow\infty} (- r^2 \partial_r e^{-\mathcal{K}/2}) = \frac{1}{4} \left(\frac{p_0}{h_0}+3 \frac{q^1}{h^1}\right)\ .
\end{align}
This illustrates that the above spacetime is supersymmetric since $M = |\mathcal{Z}|$.

A slightly more challenging example is provided if we take the supersymmetric Kerr-Newman spacetime from section 4.2 of \cite{BLS}. We will literally consider the same solution, taken in minimal supergravity with a prepotential $F = - \frac{i}{4} (X^0)^2$, such that $e^{-\mathcal{K}} = X^0 \bar{X}^0$. In oblate spheroidal coordinates (c.f.\ (59) of \cite{BLS}), the harmonic functions that give the solution are $$H_0 = 1 + \frac{m r}{r^2+\alpha^2 \cos^2 \theta}\ , \qquad H^0 = \frac{2 \alpha \cos \theta}{r^2 + \alpha^2 \cos^2 \theta}\ .$$ Solving for the vector field strengths from this, we find that $q_0 = m, p^0 = 0$. This means that
\begin{align}
    \mathcal{Z} = e^{\mathcal{K}/2} X^0 m \quad \Rightarrow \quad |\mathcal{Z}| = m\ .
\end{align}
The K\"{a}hler connection (c.f.\ \eqref{gaugedU1}) in this example is in fact non-vanishing, $A_{\theta} = \frac{1}{2} e^{\mathcal{K}/2} (H_0 \partial_{\theta} H^0 - H^0 \partial_{\theta} H_0)$. However, it goes as $r^{-2}$ as $r \rightarrow \infty$ and therefore does not contribute to the supercharge anticommutator and keeps the Minkowski asymptotics. If we further perform a redefinition $r \rightarrow r - m$, we obtain a stationary supersymmetric metric in the familiar form
\begin{align}\label{Kerr-newman}
\begin{split}
    {\rm d} s^2 &= \frac{(r-m)^2+\alpha^2 \cos^2 \theta}{r^2 + \alpha^2 \cos^2 \theta} ({\rm d} t^2 + \frac{(2 m r - m^2) \alpha \cos^2 \theta}{(r-m)^2 + \alpha^2 \cos^2 \theta} {\rm d} \varphi^2) - \frac{r^2 + \alpha^2 \cos^2 \theta}{(r-m)^2+\alpha^2} {\rm d} r^2 \\ &- (r^2 + \alpha^2 \cos^2 \theta) {\rm d} \theta^2 - (r^2 + \alpha^2 \cos^2 \theta) \frac{(r-m)^2+\alpha^2}{(r-m)^2 + \alpha^2 \cos^2 \theta} \sin^2 \theta {\rm d} \varphi^2 \ ,
\end{split}
\end{align}
which is the Kerr-Newman metric with equal mass and charge, leading to a nakedly singular rotating asymptotically flat spacetime. The mass can be again found by
\begin{align}\label{simple_mass_kerr}
    M = ... = \lim_{r\rightarrow\infty} (- r^2 \partial_r e^{-\mathcal{K}/2}) = m = |\mathcal{Z}|\ ,
\end{align}
after converting back to spherical coordinates\footnote{Eq.\ \eqref{simple_mass_kerr} holds also in the given set of Boyer-Lindquist coordinates, but in order to use \eqref{mass_Mink} one needs to first convert the relevant asymptotic quantities in spherical coordinates.}. This confirms that the Kerr-Newman metric \eqref{Kerr-newman} is supersymmetric and that the angular momentum, $J = \alpha m$, indeed does not enter in the BPS bound \eqref{bound_Mink} and remains unconstrained by supersymmetry.

 \section{AdS$_4$ asymptotics}\label{sect:AdS}
 \subsection{General analysis}
The necessary conditions for AdS$_4$ vacuum, derived in \cite{Hristov:2009uj}, are:
\begin{align}\label{AdS_asymptotic}
\begin{split}
k^i_\Lambda \bar{L}^\Lambda  =0\ ,& \qquad
\tilde k^u_\Lambda L^\Lambda =0 \\
P^x_\Lambda f_i^\Lambda =0\ ,& \qquad  \epsilon^{xyz} P^y_{\Lambda} P^z_{\Sigma} L^{\Lambda} \bar{L}^{\Sigma} = 0\ ,
\end{split}
\end{align}
with constant scalars, vanishing field strengths $F^\Lambda_{\mu \nu}=0$ and AdS$_4$ metric with cosmological constant\footnote{$\Lambda$ is the cosmological constant of pure AdS$_4$ with constant scalars. The curvature of all asymptotic AdS solutions will approach this value as $r \rightarrow \infty$. The reason for defining $g'$ is because the AdS Killing spinors explicitly contain this constant instead of the gauge coupling constant $g$.} $\Lambda \equiv -3 g'^2  = -3 g^2 P^x_{\Lambda}
P^x_{\Sigma} L^{\Lambda} \overline{L}^{\Sigma}$. \eqref{AdS_asymptotic} will have to hold at $r \rightarrow \infty$ for all asymptotically AdS spacetimes, together with the usual conditions on the metric \cite{Hristov:2011ye}. Note that we do not allow for asymptotic magnetic charge for the graviphoton, i.e.\ $P^x_{\Lambda} A^{\Lambda}_{\varphi} = 0$. Unlike in the minimal case, this does not rule out the existence of magnetic charges but only restricts them.

The last condition in \eqref{AdS_asymptotic} tells us that the $P^x_{\Lambda} L^{\Lambda}$'s are restricted in a certain way. We will assume that they are aligned in one particular direction asymptotically\footnote{$P^x \equiv P^x_{\Lambda} L^{\Lambda}$ rotates under $Sp(1) \simeq SU(2)$ and can always be put in a particular direction. This however does not mean that existing solutions in literature will automatically be written in such a way.} (direction $a$), i.e. only $P^a \equiv P^a_{\Lambda} L^{\Lambda} \neq 0$. The Majorana Killing spinors for AdS were derived in \cite{unpublished,Hristov:2011ye},
\begin{equation}\label{Killing_spinors-AdS}
    \tilde{\epsilon}^{1,2}_{AdS} = e^{\frac{i}{2} arcsinh (g' r) \gamma_1} e^{\frac{i}{2} g' t \gamma_0} e^{-\frac{1}{2} \theta \gamma_{1 2}} e^{-\frac{1}{2}
    \varphi \gamma_{2 3}}\ \tilde{\epsilon}^{1,2}_0\ ,
\end{equation}
where it was implicitly assumed that $a=2$ for the gauging in the minimal case. The end result for the supercharge anticommutator will of course not depend on which direction for the moment maps is chosen, but when $a=2$ the Killing spinors (the chiral ones can again be found using \eqref{spinor_relation}) take the simplest form. In the explicit formulas for the asymptotic charges it is clear how to leave the choice for the direction $a$ completely arbitrary. The basic anticommutator for asymptotically AdS solutions can be again derived directly using the chiral version of \eqref{Killing_spinors-AdS} in \eqref{basic_susy_anticommutator_general}. The result takes the expected form from the $OSp(2|4)$ superalgebra,
\begin{align}\label{osp_algebra}
\{ Q^{A \alpha},Q^{B \beta} \} &=  \delta^{AB}
(\hat{\gamma}^{MN}C^{-1})^{\alpha \beta} M_{MN} - \epsilon^{AB} T (C^{-1})^{\alpha\beta}\ ,
\end{align}
as discussed in detail in sections 3.1 and 4.1 of \cite{Hristov:2011ye}. Here we also require that the $U(1)$ and $Sp(1)$ gauged conections in \eqref{basic_susy_anticommutator_general} fall off fast enough as $r \rightarrow \infty$ in order to precisely recover the above expression. \eqref{osp_algebra} can be taken as a definition of asymptotically AdS spacetimes. Any spacetime, whose Dirac bracket \eqref{basic_susy_anticommutator_general} does not simplify to \eqref{osp_algebra} is therefore not asymptotically AdS. In the explicit example that follows the fall off will already be of the correct type, but in principle one needs to always make sure that the spacetime in question really is asymptotically AdS in the sense of \eqref{AdS_asymptotic} and \eqref{osp_algebra}. Each of the asymptotic charges $M_{M N}$ and $T$ can be explicitly derived, but we will again concentrate on the mass and charge in the stationary case.

 \subsection{Stationary solutions}
 Now we consider any stationary asymptotically AdS$_4$ solution (see the next subsection for an explicit example). For asymptotically AdS solutions with vanishing magnetic charge $\lim_{r\rightarrow\infty} P^x_{\Lambda} p^{\Lambda} = 0$, the supersymmetry anticommutator is\footnote{Again, the supercharges are rescaled for convenience.}
 \begin{align}\label{susy_commut_AdS}
    \{ Q^{A \alpha}, Q^{B \beta} \} = \delta^{A B} 8 \pi  ((M \gamma^0 + g' J_{i j} \gamma^{i j}) \mathcal{C}^{-1})^{\alpha \beta} - \epsilon^{A B} 8 \pi T (\mathcal{C}^{-1})^{\alpha \beta}\ ,
 \end{align}
 with\footnote{Note that the following expression includes both the gauge coupling constant $g$ and the asymptotic cosmological constant $g'$.}
 \begin{align}\label{mass_AdS}
 \begin{split}
 M &= \frac{1}{8\pi}\lim_{r\rightarrow\infty} \oint {\rm d}\Sigma_{t r} \bigg( e^t_{[ 0} e^r_1 e^{\theta}_{2]} + \sin \theta\ e^t_{[ 0} e^r_1 e^{\varphi}_{3]} \\
&+2 g g' r |P^a_{\Lambda} L^{\Lambda}|\ e^t_{[0} e^r_{1]} - \sqrt{g'^2 r^2 +1} (\omega_{\theta}^{a b} e^t_{[ 0} e^r_a e^{\theta}_{b]} + \omega_{\varphi}^{a b} e^t_{[ 0} e^r_a e^{\varphi}_{b]} ) \bigg)\ ,
 \end{split}
 \end{align}
 and
 \begin{align}
    T = \frac{1}{4\pi} \lim_{r \rightarrow \infty} \oint_{S^2}{\rm Re} \left( T^- \right) = \lim_{r \rightarrow \infty}  {\rm Re}\left( L^{\Lambda} q_{\Lambda} - M_{\Lambda} p^{\Lambda} \right)\ .
  \end{align}
 The angular momenta $J_{i j}$ remain exactly as given in App. C of \cite{Hristov:2011ye}, unaffected directly by the scalars. The BPS bound is given by
 \begin{align}\label{bound_AdS}
    M \geq |T| + g' |\vec{J}|\ .
 \end{align}
 Note that the scalars enter explicitly in the definition of the mass \eqref{mass_AdS}, unlike for the asymptotically flat solutions.

 \subsection{Static example}
 Here we will explicitly consider the static supersymmetric spacetimes with non-constant scalars constructed by Sabra in
\cite{Sabra}\footnote{These are the most general static BPS configurations that have been constructed so far in
AdS. Strictly speaking, they do not correspond to black holes but rather to naked singularities due to the
absence of an event horizon.}. Unlike in the asymptotically flat case, one cannot easily find what the mass is
just from looking at the metric.

 Briefly summarized, the solution of \cite{Sabra} is in a FI gauged supergravity with constant parameters $P^a_{\Lambda} = \xi_{\Lambda}$ and an arbitrary number of vector multiplets. The solutions are purely electric with arbitrary charges $q_{\Lambda}$. The metric and symplectic sections are
 \begin{align}\label{sabra_solution}
 \begin{split}
 {\rm d} s^2 = e^{\mathcal{K}} \left(1 + g^2 r^2 e^{-2 \mathcal{K}}\right) {\rm d} t^2 - \frac{e^{-\mathcal{K}} {\rm d} r^2}{\left(1 + g^2 r^2 e^{-2 \mathcal{K}}\right)} -e^{-\mathcal{K}} r^2 {\rm d} \Omega_2^2\ ,\\
 {\rm Im} X^{\Lambda} = 0, \qquad  2\ {\rm Im} F_{\Lambda} = H_{\Lambda} = \xi_{\Lambda} + \frac{q_{\Lambda}}{r}\ .
 \end{split}
 \end{align}
 It is immediately clear that the charge $T$ of this configuration will be
 \begin{align}\label{T}
    T = \lim_{r \rightarrow \infty}  {\rm Re}\left( L^{\Lambda} q_{\Lambda} - M_{\Lambda} p^{\Lambda} \right) = \lim_{r \rightarrow \infty} L^{\Lambda} q_{\Lambda} = e^{\mathcal{K}(\xi)/2} X^{\Lambda} (\xi) q_{\Lambda}\ ,
 \end{align}
where $\mathcal{K}(\xi), X^{\Lambda} (\xi)$ denote the corresponding asymptotic values that will only depend on the gauge parameters via the second row of \eqref{sabra_solution}. Since the solutions are supersymmetric and static ($J_{i j} = 0$) it follows that the mass takes the exact same value as the charge $T$. We can show this explicitly for any given solution.

Let us for simplicity take the prepotential $F =-2 i \sqrt{X^0 (X^1)^3}$ with electric charges $q_0,q_1$ and FI parameters $\xi_0,\xi_1$. The sections are therefore $X^0 = \frac{1}{6 \sqrt{3}} \sqrt{\frac{(H_1)^3}{H_0}}, X^1 = \frac{1}{2 \sqrt{3}} \sqrt{H_0 H_1}$ with $e^{-\mathcal{K}} = \frac{2}{3 \sqrt{3}} \sqrt{H_0 (H_1)^3}$ and $g' = \frac{2^{1/2}}{3^{3/4}} g (\xi_0 (\xi_1)^3)^{1/4}$. The asymptotic charge $T$ from \eqref{T} becomes
\begin{align}
    T = \frac{(\xi_0 (\xi_1)^3)^{1/4}}{2^{3/2} 3^{3/4}} \left(\frac{q_0}{\xi_0} + 3 \frac{q_1}{\xi_1} \right)\ .
\end{align}

In order to find the mass of this configuration we first need to perform a simple coordinate rescaling to make sure that the metric asymptotes to AdS in spherical coordinates (equivalently we could insist that $e^{-\mathcal{K}}$ asymptotes to $1$). Transforming $r \rightarrow a r, t \rightarrow  t/a$, with $a = \lim_{r \rightarrow \infty} e^{-\mathcal{K}/2} = \frac{2^{1/2}}{3^{3/4}} (\xi_0 (\xi_1)^3)^{1/4}$ we achieve
 \begin{align}\label{sabra_solution-example}
 {\rm d} s^2 =  \left(a^2 e^{\mathcal{K}} + g^2 r^2 e^{-\mathcal{K}}\right) {\rm d} t^2 - \frac{{\rm d} r^2}{\left(a^2 e^{\mathcal{K}} + g^2 r^2 e^{- \mathcal{K}}\right)} -\frac{e^{-\mathcal{K}}}{a^2} r^2 {\rm d} \Omega_2^2\ ,
 \end{align}
which exactly asymptotes to AdS with cosmological constant $- 3 g'^2$ in spherical coordinates. The functions that further define the metric now take the form $H_0 = \xi_0 + \frac{a q_0}{r}, H_1 = \xi_1 + \frac{a q_1}{r}$. The relevant spin connection components in this case are $\omega_{\theta}^{12} = \frac{\omega_{\varphi}^{13}}{\sin \theta} = \sqrt{a^2 e^{\mathcal{K}} + g^2 r^2 e^{- \mathcal{K}}} \partial_r (\frac{r e^{-\mathcal{K}/2}}{a})$. Now we can use \eqref{mass_AdS} to find the mass of this configuration:
\begin{eqnarray}
\nonumber M &= \lim_{r\rightarrow\infty} \frac{e^{-\mathcal{K}/2}}{a^2} r^2 \left( \frac{a}{r} + g g' r (\xi_0 X^0 + \xi_1 X^1) - \frac{1}{r} \sqrt{g'^2 r^2 + 1} \sqrt{a^2 e^{\mathcal{K}} + g^2 r^2 e^{-\mathcal{K}}} \partial_r (r e^{-\mathcal{K}/2}) \right) \\
 &= ... = \frac{(\xi_0 (\xi_1)^3)^{1/4}}{2^{3/2} 3^{3/4}} \left(\frac{q_0}{\xi_0} + 3 \frac{q_1}{\xi_1} \right) = T\ ,
\end{eqnarray}
as expected. This is a rather non-trivial check that \eqref{mass_AdS} gives the correct expression for the AdS mass, and therefore reproduces correctly results from holographic renormalization \cite{Kostas}. Interestingly, we note that in the process of simplifying the above formula, in ``$...$'' one finds the mass to be
\begin{align}\label{simple_mass_AdS}
    M = \lim_{r\rightarrow\infty} (- \frac{r^2}{a} \partial_r e^{-\mathcal{K}/2}) = \frac{(\xi_0 (\xi_1)^3)^{1/4}}{2^{3/2} 3^{3/4}} \left(\frac{q_0}{\xi_0} + 3 \frac{q_1}{\xi_1} \right)\ ,
\end{align}
i.e. picking the first subleading term of the K\"{a}hler potential after normalizing it to asymptote to $1$.
This simple formula turns out to give the mass for the static solutions both in Minkowski (c.f.\
\eqref{simple_mass_Mink} and \eqref{simple_mass_kerr}) and in AdS. We now turn to magnetic AdS asymptotics and
show that the same formula effectively gives the mass also for supersymmetric solutions in mAdS.

 \section{mAdS$_4$ asymptotics}\label{sect:mAdS}
 \subsection{General analysis}
 Magnetic AdS (or mAdS) was recently introduced as a concept in \cite{Hristov:2011ye}. Many features of it are similar to the purely AdS case, but due to the presence of magnetic charges mAdS preserves less supersymmetry. The asymptotic conditions on the spacetime remain as in \eqref{AdS_asymptotic} with constant scalars, only now the magnetic field strengths are $2 F^{\Lambda}_{\theta \varphi} = p^{\Lambda} \sin \theta$ under the restriction $2 g P^a_{\Lambda} p^{\Lambda} = \mp 1$ coming from Dirac quantization\footnote{Note that there is a mismatch of a factor of $2$ between the charges here and in the previous sections. It can be traced back to the different conventions used in \cite{Romans:1991nq} and \cite{Hristov:2010ri} and is compensated for in all formulas of this section.}. As before, we have the redefinition of the cosmological constant to be $\Lambda \equiv -3 g'^2$ and assume the moment map in direction $P^a$ to be non-zero.

 For $a=2$, the Killing spinors of mAdS$_4$ were given in \cite{Romans:1991nq,Hristov:2011ye}. Here we can give the projections obeyed by the chiral Killing spinors as straightforward generalization of the analysis in \cite{Hristov:2010ri}:
 \begin{equation}\label{KS-ansatz}
\varepsilon_{mAdS, A} =  e^{i \alpha}\, \epsilon_{A B} \gamma^0
\varepsilon^B_{mAdS}\ , \qquad \varepsilon_{mAdS, A} =  \pm e^{i
\alpha}\,{\sigma}^a_{A B}\,\, {\gamma}^1\, \varepsilon^B_{mAdS}\ ,
\end{equation}
where $\alpha$ is an arbitrary constant phase, and the choice of sign of the second projection corresponds to the choice of sign for the charge quantization condition. We choose to set $\alpha = 0$, which can be done without any loss of generality. However, some explicit solutions in literature might implicitly use $\alpha = \pi/2$ or other choices, which results in rotation of the symplectic sections $\{F_{\Lambda}, X^{\Lambda} \}$ by $e^{i \alpha}$ in all the equations that follow. The functional dependence of the Killing spinors can also be found in \cite{Hristov:2010ri} - it is only radial, $\sqrt{g' r + \frac{g'}{2 g^2 r}}$. This can be seen explicitly by analyzing the Killing spinor equation $\widetilde{\mathcal{D}}_{\mu} \varepsilon_A = 0$. Solving it also forces all asymptotically mAdS spacetimes to satisfy $P^a_{\Lambda} X^{\Lambda} = \pm 1, 4 g e^{\mathcal{K}} F_{\Lambda} p^{\Lambda} = \pm i$ as $r \rightarrow \infty$.

 For asymptotically mAdS solutions with non-vanishing magnetic charge, the supersymmetry anticommutator is just
 \begin{align}\label{susy_commut_mAdS}
    \{ Q^I, Q^J \} = \delta^{I J} 8 \pi M\ ,
 \end{align}
 with only two supercharge singlets as discussed in detail in section 4.2 of \cite{Hristov:2011ye}. The mass is given by explicitly plugging \eqref{KS-ansatz} in \eqref{basic_susy_anticommutator_general} for any asymptotically mAdS solution. Just as in \cite{Hristov:2011ye}, it turns out that the expression takes more convenient form if we choose an upper triangular vielbein\footnote{Note that the mass can only be defined upto an overall multiplicative constant, since one can always rescale the asymptotic Killing spinor by $k$, changing the mass by $k^2$. For Minkowski and AdS, there are already well-established conventions that fix $k$, but this is not the case for mAdS.}:
 \begin{align}\label{mass_mAdS}
 \begin{split}
M &= \frac{1}{8\pi} \lim_{r\rightarrow\infty}  \oint {\rm d}\Sigma_{t r} \left( g' r + \frac{g'}{2 g^2 r} \right)
\bigg(\mp 2\ {\rm Im}\left( L^{\Lambda} q_{\Lambda} - M_{\Lambda} p^{\Lambda} \right) \sin \theta\ e^t_{0} e^r_1 e^{\theta}_{2} e^{\varphi}_{3}  \\
&+ 2 g |P^a_{\Lambda} L^{\Lambda}|\ e^t_{0} e^r_{1} - (\omega_{\theta}^{1 2} e^t_{0} e^r_1 e^{\theta}_{2} +
\omega_{\varphi}^{1 3} e^t_{0} e^r_1 e^{\varphi}_{3} ) \bigg)\ ,
 \end{split}
 \end{align}
The BPS bound in this case is simply
 \begin{align}\label{bound_mAdS}
    M \geq 0\ .
 \end{align}
 Note that there is a crucial difference between the AdS and the mAdS masses since the scalars enter differently in the expressions, e.g.\ in the first term on the r.h.s.\ of \eqref{mass_mAdS}. We will see in the next subsection that this ultimately leads to a different notion of the mass in the two cases and that the standard holographic renormalization technique is equivalent to the mass definition \eqref{mass_AdS}, but does not reproduce correctly \eqref{mass_mAdS}. Note however that one can define another conserved charge for asymptotically mAdS spacetimes, in analogy to the central charge that appears in the Riemann AdS superalgebra \cite{Hristov:2012bd},
\begin{equation}\label{z}
Z \equiv \lim_{r \rightarrow \infty} r  (\frac{1}{2} \pm 2 g' Im (L^{\Lambda} q_{\Lambda} - M_{\Lambda} p^{\Lambda}))\ .
\end{equation}
This is always finite, since we have the identity $\lim_{r \rightarrow \infty} 2 g' Im (L^{\Lambda} q_{\Lambda} - M_{\Lambda} p^{\Lambda}) = \mp 1/2$. In this case $Z$ does not play any role in the superalgebra, but seems to be relevant when one computes masses via the holographic renormalization procedure (see more later).

 \subsection{Black hole example}
 Here we concentrate on the static supersymmetric black holes with magnetic charges, found recently by \cite{klemm-adsBH} and generalized by \cite{Dall'Agata:2010gj,Hristov:2010ri}. The theory is again FI gauged supergravity with an arbitrary number of vector multiplets and gaugings $\xi_{\Lambda}$. The magnetic charges are restricted by the equation $2 g \xi_{\Lambda} p^{\Lambda} = 1$\footnote{We just choose the positive sign here without any loss of generality.}, and the metric and scalars are given by
 \begin{align}\label{our_solution}
 \begin{split}
 {\rm d} s^2 = e^{\mathcal{K}} \left(g r+\frac{c}{2 g r} \right)^2 {\rm d}t^2 - \frac{e^{-\mathcal{K}} {\rm d}r^2}{\left(g r+\frac{c}{2 g r} \right)^2} - e^{-\mathcal{K}} r^2 {\rm d} \Omega_2^2\ ,\\
 {\rm Re} X^{\Lambda} = H^{\Lambda} = \alpha^{\Lambda} + \frac{\beta^{\Lambda}}{r}, \qquad \qquad {\rm Re} F_{\Lambda} = 0\ ,\\
 \xi_{\Lambda} \alpha^{\Lambda} = - 1\ , \qquad \xi_{\Lambda}
  \beta^{\Lambda} = 0\ , \qquad F_{\Lambda} \left( -2 g^2 r \beta^{\Lambda} + c
  \alpha^{\Lambda}+2 g p^{\Lambda}\right)= 0\ .
 \end{split}
 \end{align}
 If we evaluate the mass of this solutions from \eqref{mass_mAdS} we get the supersymmetric value $M=0$.

 To see this in some detail, let us again consider the simplest case of prepotential $F =-2 i \sqrt{X^0 (X^1)^3}$ that was also discussed carefully in section 7.1 of \cite{Hristov:2010ri}. We have $X^0 = H^0 = \alpha^0+\frac{\beta^0}{r}, X^1 = H^1 = \alpha^1+\frac{\beta^1}{r}$ and $e^{-\mathcal{K}} = 8 \sqrt{H^0 (H^1)^3}$, with
 \begin{equation}\label{constants}
\beta^0 = -\frac{\xi_1 \beta^1}{\xi_0}, \qquad \alpha^0 = -\frac{1}{4 \xi_0}, \qquad \alpha^1 = -\frac{3}{4 \xi_1}, \qquad c = 1 - \frac{32}{3} (g \xi_1 \beta^1)^2\ ,
\end{equation}
and magnetic charges
\begin{equation}\label{magn-charges}
p^0 = \frac{1}{g \xi_0} \left(\frac{1}{8}+\frac{8 (g \xi_1 \beta^1)^2}{3} \right), \quad p^1 = \frac{1}{g \xi_1} \left(\frac{3}{8}-\frac{8 (g \xi_1 \beta^1)^2}{3} \right)\ .
\end{equation}
We again need to rescale $t$ and $r$ in order to have the metric asymptote to mAdS in spherical coordinates just as above: $r \rightarrow a r, t \rightarrow  t/a$, with $a = \lim_{r \rightarrow \infty} e^{-\mathcal{K}/2} = \frac{2^{1/2}}{3^{3/4}} (\xi_0 (\xi_1)^3)^{-1/4}$ and cosmological constant coming from $g' = \frac{3^{3/4}}{2^{1/2}} (\xi_0 (\xi_1)^3)^{1/4}$. The metric is then
\begin{align}\label{rescaled_solution}
 {\rm d} s^2 = e^{\mathcal{K}} \left(g r+\frac{a^2 c}{2 g r} \right)^2 {\rm d}t^2 - \frac{e^{-\mathcal{K}} {\rm d}r^2}{\left(g r+\frac{a^2 c}{2 g r} \right)^2} - \frac{e^{-\mathcal{K}}}{a^2} r^2 {\rm d} \Omega_2^2\ ,
 \end{align}
and $H^0 = \alpha^0+\frac{a \beta^0}{r}, H^1 = \alpha^1+\frac{a \beta^1}{r}$. Evaluating \eqref{mass_mAdS} now gives
\begin{align}
M = \lim_{r\rightarrow\infty} \frac{e^{-\mathcal{K}/2}}{a^2} r^2 \left(g' r +\frac{g'}{2 g^2 r} \right) \left(g - \frac{2 a^2 e^{\mathcal{K}}}{r^2} (F_0 p^0 + F_1 p^1) - \frac{e^{\mathcal{K}/2}}{r} \left( g r+\frac{a^2 c}{2 g r}  \right) \partial_r (r e^{-\mathcal{K}/2}) \right)
 = 0\ .
\end{align}
 We are now in position to compare this result with the one obtained via the holographic renormalization techniques of \cite{Kostas,Batrachenko}. As found in section 9 of \cite{Hristov:2010ri}, the mass of the above black holes is non-vanishing if one uses the explicit formulas provided in \cite{Batrachenko} based on the procedure of holographic renormalization \cite{Kostas}. In fact these formulas give the same result as if \eqref{mass_AdS} were used, i.e.\ the holographic renormalization procedure does not consider the case of magnetic AdS asymptotics separately. More precisely, the holographically renormalized energy of asymptotically mAdS spacetimes is given by $\frac{g}{g'} M + Z$, i.e.\ one needs to combine \eqref{mass_mAdS} and \eqref{z} in a quantity that cannot be directly associated with the time-translation symmetry. 

 Remarkably, the effective formula that worked in the static cases for Minkowski and AdS (see \eqref{simple_mass_Mink} and \eqref{simple_mass_AdS}) turns out to give the correct result once again,
 \begin{align}\label{simple_mass_mAdS}
    M = \lim_{r\rightarrow\infty} (- \frac{r^2}{a} \partial_r e^{-\mathcal{K}/2}) = 0\ .
\end{align}
Although the fundamental mass formulas \eqref{mass_Mink},\eqref{mass_AdS} and \eqref{mass_mAdS} are a priori considerably different, it turns out that the corresponding supersymmetric solutions have such properties that in each case the mass reduces to exactly the same simple formula.

 \section{Final remarks}
To summarize, the main results of our work are the general mass formulas \eqref{mass_Mink}, \eqref{mass_AdS},
and \eqref{mass_mAdS} for asymptotically flat, AdS, and mAdS spacetimes, respectively. We confirmed the
well-known result \cite{hep-th/9509160} for the central charge in Minkowski, showing that the hypermultiplets do
not alter it. We also showed that supergravity does make a clear distinction between masses in AdS and mAdS. Our
analysis in AdS generalizes some previous works that did not allow for non-trivial scalars, e.g.\ \cite{AD}. The
results for asymptotically AdS solutions are in fact equivalent to performing the procedure of holographic
renormalization \cite{Kostas,Batrachenko}, i.e.\ \eqref{mass_AdS} can be directly used in AdS/CFT applications.
In the asymptotically mAdS case, to our best understanding, \eqref{mass_mAdS} is the relevant mass formula that needs to be used. Physically, this mass formula might seem a bit counter-intuitive as it allows for black hole solutions with vanishing mass. However, from the point of view of the superalgebra, the vanishing mass is the only possibility for BPS objects in mAdS. Therefore $M=0$ should not come as a surprise for the static magnetic black holes of \cite{klemm-adsBH,Hristov:2010ri}.

It is important to observe that the scalar profiles as functions of the radial coordinate enter explicitly in the mass formulas \eqref{mass_AdS} and \eqref{mass_mAdS}. Thus, the AdS and mAdS masses not only depend on the asymptotic values of the scalars, but also on how the scalars approach these values. This feature provides a new point of view towards the attractor mechanism in AdS/mAdS. It shows that scalars are much more restricted to behave in a particular way in comparison with the Minkowski case. Nevertheless, for the supersymmetric solutions it turned out that the mass can be described by the same formula in all three asymptotic vacua,
\begin{align}\label{simple_mass}
    M = \lim_{r\rightarrow\infty} (- \frac{r^2}{a} \partial_r e^{-\mathcal{K}/2})\ ,
\end{align}
where $a \equiv \lim_{r\rightarrow\infty} e^{-\mathcal{K}/2}$ is usually chosen to be $1$. This essentially means that the mass is the first subleading term of the K\"{a}hler potential expansion, no matter what the details of the solution and its asymptotics are. It will be interesting to understand the physical reasons behind this.

Finally, we note that the supercharge anticommutator \eqref{basic_susy_anticommutator_general} can also be used to describe other asymptotic vacua in gauged supergravity. Examples of potential use are in asymptotically Lifshitz spacetimes (a supersymmetric Lifshitz vacuum was found in \cite{Faedo,Halmagyi}) or in solutions with $AdS_2 \times S^2$ asymptotics.

\section*{Acknowledgements}

I would like to especially thank Chiara Toldo for initial collaboration and careful reading of the manuscript
and Stefan Vandoren for helpful discussions. I acknowledge support by the Netherlands Organization for
Scientific Research (NWO) under the VICI grant 680-47-603.

\appendix

\section{Details on $D=4$ $N=2$ gauged supergravity}\label{app:A}
Here we will give more details on the theory in consideration. Alternatively, see \cite{ABCDF} for a very detailed description. The bosonic part of the supergravity lagrangian was given in \eqref{lagr}-\eqref{pot2}. The supersymmetry variations under which the full action is invariant (upto higher order terms in fermions) are as follows. The gravitino variation is
\begin{equation}\label{susy-gravi}
\delta_\varepsilon \psi_{\mu A}=\widetilde{\mathcal{D}}_{\mu}\varepsilon_A\ ,
\end{equation}
with a supercovariant derivative $\widetilde{\mathcal{D}}$ as defined in \eqref{supercov_der}. The corresponding vielbein variation reads
\begin{equation}\label{susy-vielbein}
\delta_\varepsilon e_{\mu}^a=- i \overline{\psi}_{\mu A} \gamma^a \varepsilon^A - i \overline{\psi}_{\mu}^A \gamma^a \varepsilon_A\ .
\end{equation}
Note that $\overline{\psi}_{\mu A} \equiv i \psi^A_{\mu}{}^{\dag} \gamma_0$ in order to keep the correct chirality\footnote{We use the notation $\chi_A, \chi^A$ for positive/negative chirality spinors that are related to each other by complex conjugation.} (this holds similarly for all the conjugate (anti-) chiral spinors). In the vector multiplet sector (we will also consider the graviphoton here) we have the gaugino variation
\begin{equation}\label{susy-gluino}
\delta_\varepsilon\lambda^{iA}=i\nabla_\mu z^i
\gamma^\mu\varepsilon^A + G_{\mu\nu}^{-i}
\gamma^{\mu\nu}\epsilon^{AB}\varepsilon_B+gW^{iAB}\varepsilon_B\ ,
\end{equation}
where $\nabla_\mu z^i$ denotes the gauge covariant derivative of the complex scalars (when isometries $k^i_\Lambda$ of the K\"{a}hler manifold are being gauged), $G_{\mu\nu}^{i}$ are the field strengths of the vectors from the vector multiplets, and $W^{iAB}$ is the gaugino mass matrix,
\begin{equation}\label{defW}
W^{iAB}\equiv k^i_\Lambda {\bar L}^\Lambda \epsilon^{AB} + i
g^{i\bar \jmath}{\bar f}^\Lambda_{\bar
\jmath}P^x_\Lambda\sigma_x^{AB}\ .
\end{equation}
The mass matrix also includes the quaternionic moment maps $P^x_\Lambda$ from the hypermultiplet gauging\footnote{Note that in the absence of hypermultiplets, the quaternionic moment maps $P^x_\Lambda$ can be non-vanishing constants, called FI parameters and usually denoted with $\xi_{\Lambda}$}, together with $L^\Lambda = {\rm e}^{{\cal K}/2}X^\Lambda$ (in analogy,
$M_{\Lambda} \equiv {\rm e}^{{\cal K}/2} F_{\Lambda}$) and their derivatives $f^\Lambda_i\equiv{\rm e}^{{\cal K}/2}D_i X^\Lambda$. They are defined in terms of the holomorphic sections $X^{\Lambda}, F_{\Lambda}$ of special geometry and the K\"{a}hler potential
\begin{equation}\label{K-pot}
{\cal K}(z,\bar z)=-\ln\Big[i({\bar X}^\Lambda(\bar
z)F_\Lambda(z)-X^\Lambda(z) {\bar F}_\Lambda(\bar z))\Big]\ .
\end{equation}
Another important special K\"{a}hler quantity is the period matrix,
\begin{equation}\label{period-matrix}
{\overline {\cal N}}_{\Lambda \Sigma}\equiv \begin{pmatrix} D_iF_\Lambda \\
{\bar F}_\Lambda\end{pmatrix} \cdot {\begin{pmatrix} D_i X^\Sigma \\
{\bar X}^\Sigma\end{pmatrix}}^{-1}\ ,
\end{equation}
with ${\rm R}_{\Lambda \Sigma} \equiv {\rm Re}{\cal
N}_{\Lambda\Sigma}, {\rm I}_{\Lambda \Sigma} \equiv {\rm Im}{\cal
N}_{\Lambda\Sigma}$.
All these quantities are also explained in more details in \cite{Hristov:2009uj} where the analysis of fully supersymmetric vacua was accomplished. The bosonic susy variations in the vector multiplet sector are
\begin{equation}\label{susy-vecscalar}
\delta_\varepsilon z^i= \overline{\lambda}^{iA} \varepsilon_A\ ,
\end{equation}
and
\begin{equation}\label{susy-vectors}
\delta_\varepsilon A_{\mu}^{\Lambda}= 2 \bar{L}^{\Lambda} \overline{\psi}_{\mu A} \varepsilon_B \epsilon^{A B} + i f^\Lambda_i \overline{\lambda}^{iA} \gamma_{\mu} \varepsilon^B \epsilon_{A B} + h.c.\ .
\end{equation}
Finally, in the hypermultiplet sector, the hyperino variation is
\begin{equation}\label{susy-hyperino}
\delta_\varepsilon \zeta_\alpha = i\,
\mathcal{U}^{B\beta}_u\nabla_\mu q^u \gamma^\mu \varepsilon^A
\epsilon_{AB}\mathbb{C}_{\alpha\beta} + g N_\alpha^A\varepsilon_A\ ,
\end{equation}
with the vielbein $\mathcal{U}^{A\alpha}_u$ of the quaternionic metric $h_{u v}$, the gauge covariant derivative of the hypers $\nabla_\mu q^u$ (when gauging isometries ${\tilde
k}_\Lambda^u$ of the quaternion K\"{a}hler manifold), and the hyperino mass matrix
\begin{equation}\label{defN}
N_\alpha^A\equiv 2\,\mathcal{U}^A_{\alpha\,u}{\tilde k}^u_\Lambda
{\bar L}^\Lambda\ .
\end{equation}
The susy variation of the hypermultiplet scalars (hypers) is
\begin{equation}\label{susy-hypers}
\delta_\varepsilon q_u = \mathcal{U}^{A\alpha}_u \left(\overline{\zeta}_\alpha \varepsilon^A + \mathbb{C}^{\alpha \beta} \epsilon^{A B} \overline{\zeta}^{\beta} \varepsilon_B \right) .
\end{equation}

In order to derive the supercharge of the theory from the procedure described in section 2 of \cite{Hristov:2011ye}, we additionally need the Poisson/Dirac brackets of the fundamental fields. It will suffice to list the non-vanishing fermionic Dirac brackets that follow from the full lagrangian\footnote{The brackets for the bosonic fields can be derived directly from \eqref{lagr} if needed.} (see e.g. \cite{ABCDF}):
\begin{align}\label{Dirac_brackets}
\begin{split}
\{ \psi_{\mu A}(x), \epsilon^{0 \nu \rho \sigma} \overline{\psi}^B_{\rho}(x') \gamma_{\sigma}
 \}_{t=t'} &= \delta_{\mu}{}^{\nu} \delta_A{}^B \delta^3(\vec{x}-\vec{x'})\ ,\\
 \{ \lambda^{i}_A (x), - \frac{i}{2} g_{k \bar{\jmath}} \overline{\lambda}^{B \bar{\jmath}}(x')
\gamma_0
 \}_{t=t'} &= \delta_{A}{}^{B} \delta_k{}^i \delta^3(\vec{x}-\vec{x'})\ ,\\
 \{ \zeta_{\alpha}(x), - i \overline{\zeta}^{\beta}(x')
\gamma_0
 \}_{t=t'} &= \delta_{\alpha}{}^{\beta} \delta^3(\vec{x}-\vec{x'})\ .
\end{split}
\end{align}
The conventions about metric signatures, gamma matrices, (anti-)selfdual tensors that we use in this paper can be found in some previous papers \cite{Hristov:2009uj,arXiv:1005.3650,Hristov:2010ri}. Note in particular that we follow the conventions for $\epsilon^{\mu \nu \rho \sigma}$ of \cite{Hristov:2009uj}. Consequently, we define as a measure for the volume/surface integrals \begin{align}
{\rm d}\Sigma_\mu=\frac{1}{6}\epsilon_{\mu\nu\rho\sigma}\,{\rm d}x^\nu \wedge {\rm d}x^\rho\wedge {\rm d}x^\sigma\ ,\qquad
{\rm d}\Sigma_{\mu\nu}=\frac{1}{2}\epsilon_{\mu\nu\rho\sigma}\,{\rm d}x^\rho\wedge {\rm d}x^\sigma\ ,
\end{align}
which are defined differently in \cite{Hristov:2011ye}.

\section{Supersymmetry charge}\label{app:B}
From the susy variations one can fix uniquely the supersymmetry charge $\mathcal{Q}$ by the requirement that
\begin{equation}\label{susy_variation_from_Q}   \delta_{\epsilon} \phi = \{\mathcal{Q},\phi \}, \end{equation}
for all fundamental fields (here denoted by $\phi$) in the theory. From \eqref{susy-gravi}-\eqref{susy-hypers}, together with the Dirac brackets \eqref{Dirac_brackets}, one finds
\begin{align}\label{susy_charge_beg}
\begin{split}
    \mathcal{Q} = \int_V {\rm d} \Sigma_{\mu} &[ \epsilon^{\mu \nu \rho \sigma} \overline{\psi}_{\nu}^A \gamma_{\rho} \tilde{\mathcal{D}}_{\sigma} \epsilon_A + h.c. \\
    &- i g_{i \bar{\jmath}} \overline{\lambda}^{\bar{\jmath}}_A  \gamma^{\mu} (i\nabla_\nu z^i
\gamma^\nu\varepsilon^A + G_{\nu\rho}^{-i}
\gamma^{\nu\rho}\epsilon^{AB}\varepsilon_B+gW^{iAB}\varepsilon_B\ ) + h.c. \\
&- i \overline{\zeta}^{\alpha} \gamma^{\mu} (i\,
\mathcal{U}^{B\beta}_u\nabla_\nu q^u \gamma^\nu \varepsilon^A
\epsilon_{AB}\mathbb{C}_{\alpha\beta} + g N_\alpha^A\varepsilon_A\ ) + h.c. ]\ ,
\end{split}
\end{align}
up to higher order in fermions. The expression for the supercharge simplifies considerably when evaluated on shell, due to the very suggestive form of the equations of motion of the gravitinos:
\begin{align}\label{eom_gravitino}
\begin{split}
\epsilon^{\mu \nu \rho \sigma} \gamma_{\nu} \tilde{\mathcal{D}}_{\rho} \psi_{\sigma A} &= g_{i \bar{\jmath}} (\nabla^\mu \bar{z}^{\bar{\jmath}} \lambda^{i}_A-\nabla_{\nu} z^i \gamma^{\mu \nu} \lambda^{\bar{\jmath}}_A) - i g_{i \bar{\jmath}} ( G_{\mu \nu}^{+ \bar{\jmath}}
\gamma^{\nu}\epsilon_{AB} \lambda^{i B} + gW^{i}_{AB} \gamma^{\mu} \lambda^{\bar{\jmath} B} ) \\
& - (\mathcal{U}^{B\beta}_u\nabla^{\mu}q^u \epsilon_{AB}\mathbb{C}_{\alpha\beta} - \mathcal{U}^{B\beta}_u\nabla_{\nu}q^u \gamma^{\mu \nu} \epsilon_{AB}\mathbb{C}_{\alpha\beta} + i g N_{\alpha A} \gamma^{\mu}) \zeta^{\alpha}\ .
\end{split}
\end{align}
After performing a partial integration of the first term on the r.h.s.\ of \eqref{susy_charge_beg} and using \eqref{eom_gravitino}, the supercharge becomes a surface integral:
\begin{align}\label{susy_charge}
\mathcal{Q} \stackrel{e.o.m.}{=} \oint_{\partial V} {\rm d} \Sigma_{\mu \nu} \epsilon^{\mu \nu \rho \sigma} \left( \overline{\psi}_{\sigma}^A \gamma_{\rho} \varepsilon_A - \overline{\psi}_{\sigma A} \gamma_{\rho} \varepsilon^A \right)\ ,
\end{align}
similarly to (2.26) in \cite{Hristov:2011ye} in the minimal case.

\end{document}